\documentclass[conference,letterpaper]{IEEEtran}

\usepackage[utf8]{inputenc} 
\usepackage[T1]{fontenc}    
\usepackage{amsfonts}       
\usepackage{nicefrac}       
\usepackage[cmex10]{amsmath} 
\usepackage[capitalize,noabbrev]{cleveref}
\usepackage{tikz}
\usepackage{pgfplots}
\usepackage{pgfplotstable} 
\pgfplotsset{compat=1.17}
\usepackage[backend=biber,style=ieee,giveninits]{biblatex}
\usepackage[ruled]{algorithm2e}
\usepackage{MnSymbol}%

\usepackage{booktabs} 
\usepackage{csquotes}
\usepackage{subcaption}

\newtheorem{theorem}{Theorem}
\newtheorem{prop}{Proposition}
\newtheorem{property}{Property}

\newcommand{\bigO}{\mathcal{O}}
\newcommand{\FBM}{FBM}
\newcommand{\DAAS}{DAAS}
\newcommand{\etal}{\emph{et al.}}
\newcommand{\DTV}{D_{TV}}
\newcommand{\DKL}{D_{KL}}
\newcommand{\DWO}{D_{W_1}}

\definecolor{myplotgreen}{RGB}{0, 158, 115} 
\definecolor{myplotred}{RGB}{213, 94, 0}   
\definecolor{myplotblue}{RGB}{0, 114, 178}

\begin{document}
\newcounter{colcount}

\pgfplotstableread[header=false, col sep=comma]{images/plot_data/approximation-convergence-uniform.csv}\datatableUniform
\pgfplotstableread[header=false, col sep=comma]{images/plot_data/approximation-convergence-triangular.csv}\datatableTriangular
\pgfplotstableread[header=false, col sep=comma]{images/plot_data/approximation-convergence-quadratic.csv}\datatableQuadratic

\pgfplotstablegetrowsof{\datatableUniform} \pgfmathsetmacro{\numrowsUniform}{\pgfplotsretval}
\pgfplotstablegetrowsof{\datatableTriangular} \pgfmathsetmacro{\numrowsTriangular}{\pgfplotsretval}
\pgfplotstablegetrowsof{\datatableQuadratic} \pgfmathsetmacro{\numrowsQuadratic}{\pgfplotsretval}

\pagestyle{plain}

\title{Discretized Approximate Ancestral Sampling}

\author{%
  \IEEEauthorblockN{Alfredo De la Fuente}
  \IEEEauthorblockA{Google \\
                    alfredodlf@google.com}
  \and
  \IEEEauthorblockN{Saurabh Singh}
  \IEEEauthorblockA{Google DeepMind \\
                    saurabhsingh@google.com}
  \and
  \IEEEauthorblockN{Jona Ballé}
  \IEEEauthorblockA{New York University \\
                    jona.balle@nyu.edu}
}

\maketitle

\begin{abstract}
The Fourier Basis Density Model (\FBM)~\cite{delafuente2024fourierbasisdensitymodel} was recently introduced as a flexible probability model for band-limited distributions, i.e. ones which are smooth in the sense of having a characteristic function with limited support around the origin. Its density and cumulative distribution functions can be efficiently evaluated and trained with stochastic optimization methods, which makes the model suitable for deep learning applications. However, the model lacked support for sampling. Here, we introduce a method inspired by discretization--interpolation methods common in Digital Signal Processing, which directly take advantage of the band-limited property. We review mathematical properties of the \FBM{}, and prove quality bounds of the sampled distribution in terms of the total variation (TV) and Wasserstein--1 divergences from the model. These bounds can be used to inform the choice of hyperparameters to reach any desired sample quality. We discuss these results in comparison to a variety of other sampling techniques, highlighting tradeoffs between computational complexity and sampling quality.
\end{abstract}

\section{Introduction}

Probability density models are fundamental tools across a multitude of disciplines, enabling tasks ranging from statistical inference and anomaly detection to generative modeling and reinforcement learning. Fitting them to data can provide a means to understand the underlying distribution, and generate new samples consistent with the observed patterns. The Fourier Basis Density Model (\FBM) \cite{delafuente2024fourierbasisdensitymodel} is a simple yet powerful parametric density modeling approach. It represents the density as a truncated Fourier series, which imposes smoothness on the probability density function (PDF), but allows arbitrarily extending the number of parameters to capture increasingly non-smooth densities. The \FBM{} admits efficient evaluation of the PDF as well as the cumulative distribution function (CDF), and can be fitted using first-order optimization methods such as stochastic gradient descent. However, \cite{delafuente2024fourierbasisdensitymodel} did not introduce a method for efficiently sampling from the distribution, which limits its practical utility.

The term \emph{sampling} is used differently in statistics and signal processing. We assume the reader is familiar with the former meaning of the term, which we use throughout this paper. In signal processing, on the other hand, a continuous-time signal $s(t)$, $t \in \mathbb{R}$, is \emph{sampled} with period $T > 0$ to yield the discrete-time signal $s[n] = s(nT)$, $n \in \mathbb{Z}$.
Typically, the operation is expressed as a multiplication with the Dirac comb $\sum_{n=-\infty}^{\infty} \delta(t - nT)$.
The aim is to recover $s(t)$ from some given $s[n]$, where the reconstructed signal is generally given by a convolution with an interpolation filter $w(t)$:
\begin{equation}
\hat s(t) = \sum_{n=-\infty}^{\infty} s[n] \, w(t/T-n).
\label{eq:dsp_reconstruction}
\end{equation}
To avoid confusion, we refer to this concept of sampling as \emph{discretization}. The Nyquist--Shannon theorem \cite{shannon1949communication} states that, if $s(t)$ is band-limited and $w(t)$ is the $\mathrm{sinc}$ function, $s(t)$ can be perfectly reconstructed, i.e., $\hat s(t) = s(t)$ for all $t$.

Inspired by this, and the fact that the PDF $p(x)$ of the \FBM{} is band-limited by definition, we propose a sampling method based on discretizing it. This yields a discrete probability mass function $p[k]$, which is easy to sample from. To obtain approximate samples from $p(x)$, we add i.i.d. random noise samples from a density $w(x)$ to samples from $p[k]$, which \enquote{interpolates} the sample distribution, in direct analogy to \eqref{eq:dsp_reconstruction}. We also consider an extension of this approach based on Markov-chain Monte Carlo (MCMC) methods. As this is an example of \emph{ancestral sampling} (with $p[k]$ as the \emph{ancestor} distribution), we call our method Discretized Approximate Ancestral Sampling (\DAAS).

We begin by reviewing several standard sampling methods in the next section, followed by a review of the \FBM{} in \cref{sec:fbm}. We introduce \DAAS{} in \cref{sec:daas} and prove bounds on the deviation of the distribution of the samples vs. the model, which we test empirically in the subsequent section. The paper concludes with \cref{sec:conclusion}.

\section{Related work}

Generally, the problem of sampling from a probability distribution is ubiquitous within numerous scientific disciplines. The goal is to generate a set of samples with an empirical density $q(x)$ that closely approximates the target (model) density $p(x)$. Computationally, sampling methods are often constructed by transforming samples from simpler distributions, for example a uniform distribution, which in turn can be more directly obtained from pseudo-random number generators (PRNGs). However, not all density models admit such straight-forward constructions, including the \FBM.

Here, we review several standard sampling methods, which unfortunately all turn out to have drawbacks when applied to the \FBM{}. We revisit some of them in \cref{sec:experiments}, in the context of evaluating our method. 

\textbf{Inverse Transform Sampling.}  This method, applicable when the CDF, $P(x) = \int_{-\infty}^{x} p(t) dt$, is known and invertible, is based on the probability integral transform.  A uniformly distributed random variable $U \sim \mathcal U(0, 1)$ is generated, and the sample is obtained as $X = P^{-1}(U)$. Unfortunately, the inverse of the CDF, $P^{-1}(x)$, is not available in closed form for the \FBM{}. We could rely on numerical methods to approximate the inverse of cumulative distribution \cite{olver2013fastinversetransformsampling, giles2023approximating, Giles_2023}, however this does not exploit the \FBM{} properties and can lead to numerical instabilities. 

\textbf{Rejection Sampling.} This method relies on a \enquote{proposal} or \enquote{envelope} distribution $e(x)$, from which we can easily sample, and a constant $M$ such that $M e(x) \geq p(x)$ for all~$x$. A sample $X$ is drawn from $e(x)$, and a uniform random number $U \sim \mathcal U(0, 1)$ is generated. The sample is accepted if $U \leq p(X) / (M e(X))$; otherwise, it is rejected, and the process is repeated \cite{gentle2003random,givens2012computational}.
One advantage of the method is that it does not require the CDF or its inverse while generating unbiased i.i.d. samples. However, the main disadvantage is that it is generally difficult to choose $e(x)$ and $M$ to create a tight envelope. With a loose envelope function, the method can be computationally inefficient due to a low acceptance rate.

\textbf{Langevin Dynamics.} Markov Chain Monte Carlo Methods (MCMC) construct a Markov chain whose stationary distribution is the target distribution $p(x)$. A subset of these methods use Langevin Dynamics \cite{rossky1978brownian} to leverage the gradient information of the log density of $p(x)$ (score function) in order to guide the proposal distribution towards the target distribution more efficiently. The proposal is based on a discretized Langevin diffusion process $X_{t+1} = X_t + \epsilon_t \nabla \log p(X) + \sqrt{2\epsilon_t} Z$, where $\epsilon_t$ is a step size and $Z$ is a standard normal random variable. This sampling method is referred to as Unadjusted Langevin Algorithm (ULA). An additional improvement can be obtained by using the Metropolis--Hastings criterion to accept or reject each sample given by the ULA proposal, which corresponds to the Metropolis-adjusted Langevin algorithm (MALA) \cite{roberts2002langevin, xifara2014langevin}. Both models can be shown to converge to the target distribution given sufficient iterations.

The method presented here is partially inspired by the thesis project~\cite{Olofsson1776777}, which to the best of our knowledge first formulated the idea of sampling from a discretization of a circular, band-limited density. However, \cite{Olofsson1776777} focuses on achieving precise samples via Féjer interpolation. Since, sampling from Féjer kernels is in itself difficult, \cite{Olofsson1776777} approximates its lobes numerically and employs rejection sampling. In contrast, we focus on examining the quality of efficient approximations using simple kernels, and introduce MCMC methods as a refinement.

\section{Fourier Basis Density Model}
\label{sec:fbm}

The \FBM{} is essentially a probability distribution on a circle. For convenience, we parameterize it in terms of an angle $x \in [-1, 1)$.
We consider the PDF $p(x) \equiv f(x)/Z$, where $f(x)$ is a periodic, real-valued, non-negative function and $Z = \int_{-1}^{1}f(x) \, dx$ is the normalization constant. We define $f(x)$ in terms of its truncated Fourier expansion with $N$ frequency terms, \emph{periodic} with period~2:

\begin{equation}
\label{eq:fourier_series}
f(x) = \sum_{n=-N}^{N} c_n \, \exp(\pi inx),
\end{equation}
where $i\equiv \sqrt{-1}$ is the imaginary unit and $c_n \in \mathbb{C}$ for $n \in \{-N, \ldots, N\}$. Conversely, we can write the coefficients as
\begin{equation}
\label{eq:coefficients}
c_n = \frac 1 2 \int_{-1}^{1} f(x) \exp(-\pi inx) \, dx.
\end{equation}
$f(x)$ is \emph{real-valued} if and only if the coefficients follow the symmetry $c_{-n} = c_n^\ast$ for all $n$, where $\ast$ denotes the complex conjugate. (Note this implies $c_0 \in \mathbb{R}$.) Consequently, the coefficients with $n<0$ are redundant and need not be considered model parameters -- i.e., we can write:
\begin{align}
\label{truncatedeq}
f(x) &= c_0 + \sum_{n=1}^N \bigl( c_n \exp(\pi inx) + c_n^\ast \exp(-\pi inx) \bigr) \\
&= c_0 + 2\sum_{n=1}^N  \Re \bigl\{ c_n \exp(\pi i n x) \bigr\}
\end{align}
To ensure that $f(x)$ is \emph{non-negative},
\cite{delafuente2024fourierbasisdensitymodel} parameterizes $c_n$ as the autocorrelation of a sequence $\{a_k \in \mathbb C \}_{k=0}^N$:
\begin{equation}
c_n = \sum_{k=0}^{N-n} a_k a_{k+n}^\ast, \quad n \in \{0, \dots, N \},
\label{eq:posdef_c}
\end{equation}
which implies, by the Wiener--Khinchin theorem, that ${c_n}$ is a positive semi-definite sequence. (In particular, $c_0 = \sum_{k=0}^N |a_k|^2 \ge 0$.) Furthermore, Herglotz's theorem \cite{brockwell2013time} states that a function $f(x)$ is non-negative if and only if its Fourier coefficients ${c_n}$ are positive semi-definite. Thus, the parameterization ensures that for any choice of $a_k$, $f(x)$ is indeed non-negative. For a concise proof, see Appendix~\ref{sec:prop1_proof_app} of the pre-print of this paper, available on arXiv.

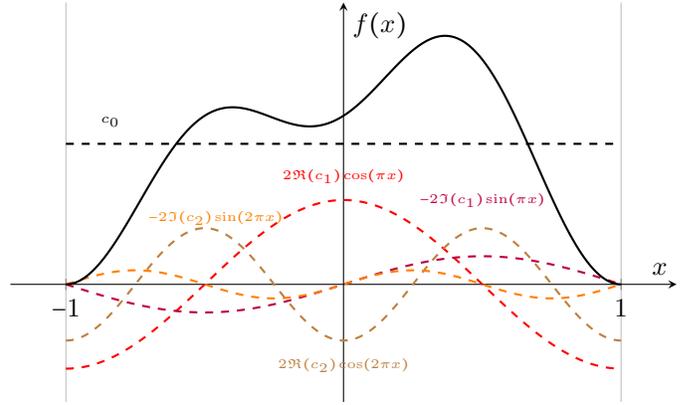
\begin{figure}
\begin{tikzpicture}
    \begin{axis}[
        width=\linewidth, height=.6\linewidth,
        scale only axis,
        domain=-1:1, samples=200,
        xlabel={$x$}, ylabel={$f(x)$},
        axis x line=middle, axis y line=middle,
        xtick={-1, 0, 1}, ytick=\empty,
        enlargelimits=true,
        grid=both,
        legend pos=north east,
        every axis plot/.append style={thick},
    ]
    \addplot[red, dashed] {0.3*cos(deg(pi*x))} node[pos=0.5, above, yshift=2.5pt] {\tiny $2\Re (c_1) \cos(\pi x)$};
    \addplot[brown, dashed] {-0.2*cos(2*deg(pi*x))} node[pos=0.5, below, yshift=-3pt] {\tiny $2\Re(c_2) \cos(2\pi x)$};
    \addplot[purple, dashed] {0.1*sin(deg(pi*x))} node[pos=0.75, above, yshift=15pt] {\tiny $ -2\Im (c_1) \sin(\pi x)$};
    \addplot[orange, dashed] {0.05*sin(2*deg(pi*x))} node[pos=0.27, above, yshift=20pt] {\tiny $-2\Im (c_2) \sin(2\pi x)$};
    \addplot[black, dashed] {0.5} node[pos=0.08, above, yshift=3pt] {\tiny $c_0$};
    \addplot[black, thick, smooth] 
        {0.5 + 0.3*cos(deg(pi*x)) - 0.2*cos(2*deg(pi*x)) + 0.1*sin(deg(pi*x)) + 0.05*sin(2*deg(pi*x))};
    \end{axis}
\end{tikzpicture}
\caption{Illustration of a band limited Fourier series (\cref{eq:fourier_series}) with only two frequency terms approximating a circular probability density within the range $[-1, 1)$.}
\label{fig:fbm_viz}
\end{figure}

The normalization constant works out to be
\begin{equation}
Z = \int_{-1}^1 f(x) dx = 2c_0.
\end{equation}
With this, the PDF can be compactly written as:
\begin{equation}
p(x) = \frac{1}{2} + \sum_{n=1}^N \Re \left\{ \frac{c_n}{c_0} \exp(\pi i n x) \right\}, \quad x \in [-1, 1),
\label{eq:model_p}
\end{equation}
with coefficients $c_n$ given by \eqref{eq:posdef_c}.
We visualize the Fourier representation of an example density in \cref{fig:fbm_viz}.

To extend the \FBM{} from the circle to the entire real line, \cite{delafuente2024fourierbasisdensitymodel} propose a change of variables using the mapping $g : (-1, 1)  \rightarrow \mathbb{R}$, which is parameterized by a scaling $s$ and an offset $t$ as follows:
\begin{equation}
g(x; s, t) = s \cdot \tanh^{-1}(x) + t = \frac s 2 \ln \left( \frac{1+x}{1-x} \right) + t.
\label{eq:transformation}
\end{equation}
Note that producing a sample of this expanded model on $\mathbb R$ is simple: Given a sample from $p(x)$, transform the sample using $g$. Hence, we can focus on obtaining a sample from the circular density in this paper.

\section{Discretized Approximate Ancestral Sampling}
\label{sec:daas}

\begin{algorithm}[t]
\KwIn{\FBM{} density $p(x)$, interpolating density $w(x)$, number of ancestors $K$, number of samples $S$.}
\KwOut{Samples from $q(x) \approx p(x)$.}
p $\leftarrow$ array($K$)

samples $\leftarrow$ array($S$)

\For {$k \leftarrow 0$ \KwTo $K-1$} {
p[$k$] $\leftarrow 2/K \cdot p(-1 + 2/K \cdot k)$
}

\For {$i \leftarrow 0$ \KwTo $S-1$} {
draw $n \sim$ p[$k$] \tcc{$n \in \{0, \ldots, K-1\}$}

draw $u \sim w$

$x \leftarrow 2/K \cdot (n+u)$

$x \leftarrow x \mod 2 - 1$ \tcc{limit to $[-1, 1)$}

samples[$i$] $\leftarrow x$
}
\Return{} samples
\caption{Discrete Approximate Ancestral Sampling}
\label{alg:daas}
\end{algorithm}

To sample from a fitted \FBM{} model $p(x)$, we propose a two-step approach, with an optional third step for refinement.

\begin{figure*}
\begin{subfigure}{0.49\textwidth}
\centering
\includegraphics[width=\textwidth]{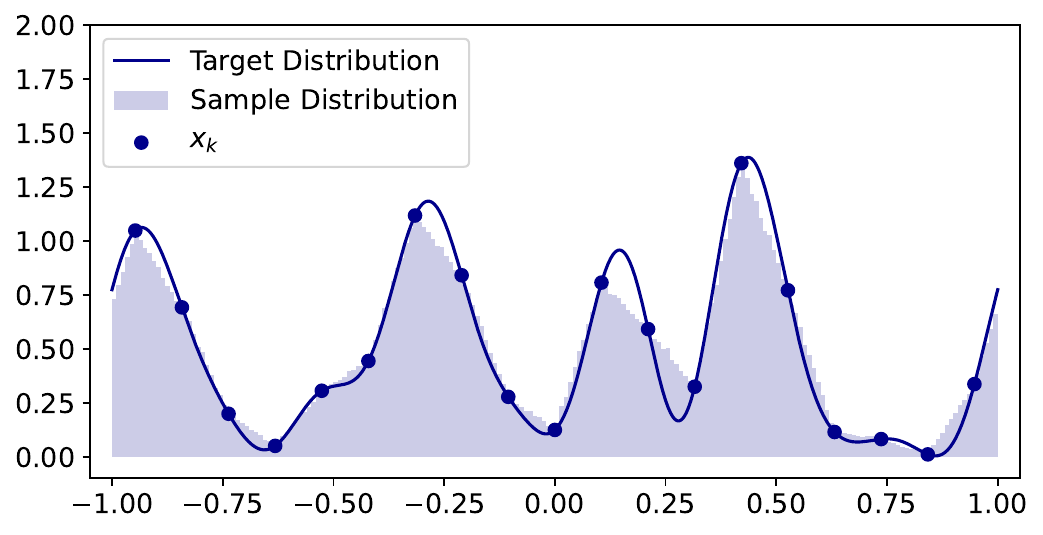}
\caption{$K=20$}
\end{subfigure}\hfill%
\begin{subfigure}{0.49\textwidth}
\centering
\includegraphics[width=\textwidth]{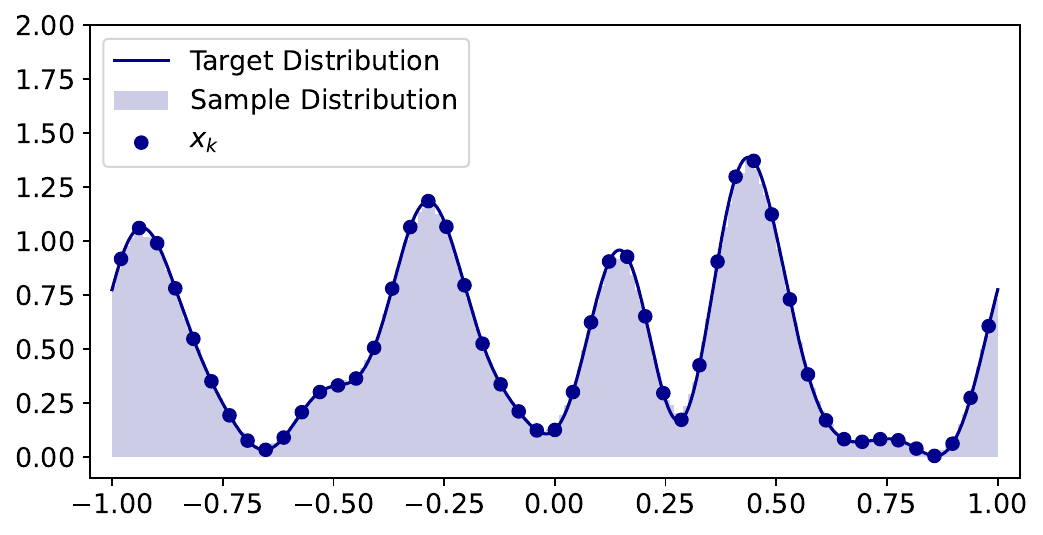}
\caption{$K=50$}
\label{fig:empirical-distributions-2b}
\end{subfigure}
\caption{Visual comparison of histograms obtained from Algorithm \ref{alg:daas} with the triangle kernel $w_1$ and varying $K =\{20, 50\}$ for an arbitrary multi-modal \FBM{} density with $10$ frequency terms. In the case of the Nyquist rate ($K=20$), the histogram clearly illustrates the piecewise linear nature of $q(x)$ (light blue fill). As we increase $K$ to 50, $q(x)$ approximates the target distribution $p(x)$ more accurately (solid blue line).}
\label{fig:empirical-distributions-2}
\end{figure*}

\subsection{Step 1: sampling from the ancestor}
First, we discretize $p(x)$ by evaluating it at $K \ge 2N$ distinct locations:
\begin{equation}
p[k] = 2/K \cdot p(x_k)
\label{eq:discretized_p}
\end{equation}
with
\begin{equation}
x_k = -1 + 2/K \cdot k, \quad k \in \{0, \ldots, K-1\}
\label{eq:ancestor_points}
\end{equation}
The set of values $\{p[k], k = 0, \ldots, K-1\}$ sum up to 1 due to the following lemma, and hence constitute a valid discrete probability distribution, which we call the \emph{ancestor distribution}.

\begin{prop}[A.6, Gillman \etal \cite{gillman2024fourierheadhelpinglarge}]
\label{prop:average_value}
Let $x_k = -1 + \frac{2k}{K}$ for $k=0, \dots, K-1$ be $K > 2N$ equally spaced points in the interval $[-1, 1)$. Then, $\sum_{k=0}^{K-1} p(x_k) = \frac{K}{2}$. For a proof, see Appendix~\ref{sec:prop5_proof_app} of the pre-print.
\end{prop}

To start the sampling procedure, we draw a sample $n \sim p[k]$.

\subsection{Step 2: sampling from approximate conditional}
The purpose of the second step is to draw a conditional sample $w(x \mid n)$ such that the resulting marginal distribution
\begin{equation}
\label{eq:q_definition}
q(x) = \sum_{k=0}^{K-1} w(x \mid k) \, p[k]
\end{equation}
approximates $p(x)$ as well as possible.

In analogy with \eqref{eq:dsp_reconstruction}, we choose $w(x \mid n)$ to be an inter\-polation filter shifted to the location of $n$ and scaled to the discretization step size:
\begin{equation}
w(x \mid n) = \tfrac K 2 w\bigl(\tfrac K 2 (x - x_n)\bigr).
\end{equation}
Although the Nyquist--Shannon theorem calls for $w(x) = \mathrm{sinc}(x)$, practical signal processing applications typically require the interpolation filter to have finite support, which rules out the $\mathrm{sinc}$ function. Here, the requirements for the density $w(x)$ are different: first and foremost, it must be non-negative and normalized in order for it to be a valid density. Hence, the $\mathrm{sinc}$ function is still inadmissible. However, $w(x)$ can generally have infinite support. For example, it could be a normal distribution.

Among many possible options, we find that cardinal B-splines are a good choice. To make $q(x)$ a B-spline interpolation of degree $D$, we can use an interpolating density $w_D(x)$ given by $D$ convolutions of a uniform density:
\begin{align}
w_0(x) &= \mathcal U\bigl(x \big\mid -\tfrac 1 2, \tfrac 1 2\bigr), \\
w_D(x) &= \bigl(w_{D-1} \ast w_0\bigr)(x).
\end{align}
This is attractive, as generating a sample from $w_D(x)$ is simple: We only need to draw $D+1$ samples from a uniform distribution, and add them together. See \cref{fig:bsplines} in the pre-print for a qualitative comparison. For example, we could choose $w_1(x)$, also called a \emph{triangle} or \emph{tent} function:
\begin{equation}
w_1(x) = \max \bigl(0, 1 - |x| \bigr).
\label{eq:triangle}
\end{equation}
Sampling from this density only requires adding two i.i.d. uniform samples, and makes $q(x)$ a linear interpolation, due to the following proposition.

\begin{prop}
\label{prop:piecewise_linear_approximation}
Let $p(x)$ be an arbitrary periodic \FBM{} and $q(x)$ a compound distribution constructed as follows:
\begin{equation}
\label{eq:q_as_mixture}
q(x) = \sum_{k=0}^{K-1} \tfrac K 2 w_1\bigl(\tfrac K 2 (x - x_k)\bigr) \, p[k],
\end{equation}
where $w_1(x)$ is the triangular kernel given by \eqref{eq:triangle}, appropriately wrapping around at the boundaries of the domain $[-1, 1)$. Then, $q(x)$ is a piecewise linear interpolation of the original distribution $p(x)$ evaluated at $x_k$. For a proof, see Appendix~\ref{sec:prop9_proof_app} of the pre-print.
\end{prop}

We characterize the properties of the linear interpolation in detail in \cref{sec:theory}. Since the conditions of the Nyquist--Shannon theorem are not satisfied, $q(x)$ will differ from $p(x)$ in general. However, the error can be controlled by the number of discretization steps K.

\emph{Step 1} and \emph{Step 2} are summarized in Algorithm~\ref{alg:daas}.

\subsection{Step 3 (optional): refinement using MCMC}
We propose to use the result of \emph{Step 2} as the initial distribution of an iterative Markov-chain Monte Carlo (MCMC) method for continuous random variables and run the chain for $T$ steps to further refine the samples. This method exploits the effect of the initial distribution on the convergence of the MCMC chains -- the closer the initial distribution to the target distribution, the faster the convergence -- and enjoys the asymptotic convergence guarantees of the MCMC methods as $T \rightarrow \infty$. We explore Unadjusted Langevin Algorithm (ULA) and Metropolis-adjusted Langevin algorithm (MALA)~\cite{roberts2002langevin, xifara2014langevin} as possible refinement methods for Algorithm~\ref{alg:daas}. ULA and MALA are summarized in Appendix \ref{sec:ula_mala_details} of the pre-print. In order to apply these algorithms to circular distributions such as the FBM, we need to wrap the sample back to the circle at each iteration.

\subsection{Computational complexity}

\subsubsection*{Sampling ancestors}
Sampling from the discrete distribution $p[k]$ can be done efficiently by using Alias Sampling \cite{walker1977efficient}. It requires $\bigO(K)$ setup time for constructing tables, after which each sample is obtained in $\bigO(1)$ time. Thus, to obtain $S$ samples it takes $\bigO(S)$ time whenever $S\gg K$. 

\subsubsection*{Evaluation of \FBM{}}
The Fast Fourier Transform (FFT) can be used to evaluate the truncated Fourier series with $N$ terms on $K$ points in $\bigO(K \log K)$ time, instead of the naive $\bigO(K N)$ approach.

\subsubsection*{Algorithm~\ref{alg:daas} with triangular noise ($w_1$)}
First the \FBM{} is evaluated at $K$ equally spaced points in $\bigO(K\log K)$. Next $S$ samples are drawn from ancestor distribution in $\bigO(S)$. Finally, triangular noise is added to each sample to obtain the approximate target samples in $\bigO(S)$. Thus, the final algorithmic complexity is $\bigO(S + K \log K )$. In particular, when $S\gg K$, our proposed Algorithm (\ref{alg:daas}) with $w_1$ produces approximate samples in time linearly proportional to the number of samples. 

\subsubsection*{Algorithm~\ref{alg:daas} with ULA/MALA}
From the previous section, samples from the initial distribution for the first Langevin iteration are obtained in $\bigO(S)$. For both ULA and MALA, the score function is evaluated at each iteration for $S$ samples in $\bigO(S\log S)$. Therefore, for $T$ iterations, the worst-case time complexity is $\bigO(T S\log S)$. For MALA, the additional accept--reject calculations only affect the constant term.

\subsection{Bounds for the approximation error}
\label{sec:theory}
For a given \FBM{} $p(x)$, let $q(x)$ be as defined in \eqref{eq:q_as_mixture}.
Then the following result bounds the divergence between the two distributions $p(x)$ and $q(x)$.
\begin{theorem}
\label{thm:divergence_bounds}
With the same setting as \cref{prop:piecewise_linear_approximation},
there exist constants $C_1 > 0, C_2 > 0$ such that,
\begin{enumerate}
\item the Total Variation Divergence ($\DTV$) is bounded by
\begin{align}
\label{eq:tvd_bound}
    \DTV(p,q) \leq  \frac{C_1}{K^2},
\end{align}
\item the Wasserstein Divergence ($\DWO$) is bounded by
\begin{align}
\label{eq:wasser_bound}
    \DWO(p,q) \leq  \frac{C_2}{K^2}.
\end{align}
\end{enumerate}
\end{theorem}
For a proof, refer to Appendix~\ref{sec:thm1_proof_app} in the pre-print. The theorem implies that for any $p(x)$, we can obtain samples of an arbitrary accuracy by choosing a large enough number of discretization points $K$.

\section{Experimental evaluation}
\label{sec:experiments}

\begin{figure*}
\centering
\includegraphics[width=0.48\linewidth, height=4.5cm]{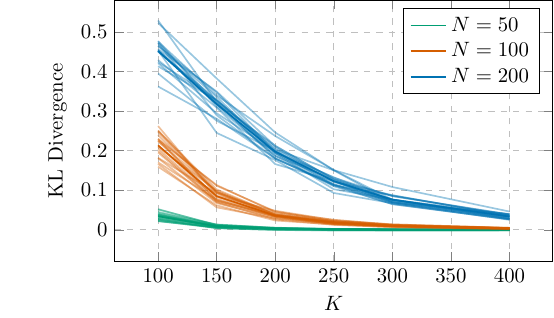}
\includegraphics[width=0.48\linewidth, height=4.65cm]{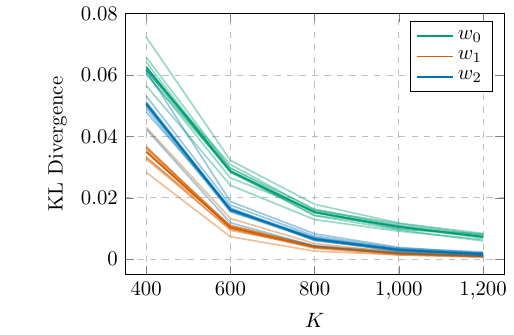}
\caption{Visualization of $\DKL$ decreasing w.r.t $K$ calculated between the unbiased samples from the target distribution via rejection sampling and the approximate samples obtained from our algorithm, considering $10$ randomly initialized FBMs. \textbf{Left: } We consider different number of frequencies $N=\{50, 100, 200\}$ for \FBM{} initializations, and observe the same trend as $K$ grows. \textbf{Right: } We explore different B-spline kernels $w_D$ for $N=50$: the linear interpolation (triangular kernel $w_1$) performs significantly better empirically than the uniform ($w_0$) or piecewise quadratic spline ($w_2$).} 
\label{fig:approximation-convergence}
\end{figure*}

\begin{figure}
\centering
\begin{tikzpicture}
\begin{axis}[
    xlabel=$T$,
    ylabel={$W_1 (\times 10^{-3})$},
    legend pos=north east,
    xmajorgrids=true,
    ymajorgrids=true,
    grid style=dashed,
    height=4.5cm,
    width=0.5\textwidth,
    xtick={0,1,2,3,4,5},
    xticklabels={0,1,5,20,100,500},
    ytick={0.5, 1.0, 1.5, 2.0},
    error bars/y dir=both,
    error bars/y explicit,
    legend pos=north east,
    legend cell align={left},
    xmin=-0.25,
    xmax=5.25,
    ymin=0.75,
    ymax=1.77,
]
\addplot[color=myplotred,mark=*,very thick] coordinates {
    (0, 1.675)
    (1, 1.655)
    (2, 1.554) 
    (3, 1.287) 
    (4, 1.053) 
    (5, 0.944) 
};
\addlegendentry{ULA}
\addplot[color=myplotgreen,mark=*,very thick] coordinates {
    (0, 1.675)
    (1, 1.432)
    (2, 1.185) 
    (3, 1.066) 
    (4, 1.005) 
    (5, 0.874) 
};
\addlegendentry{MALA}
\addplot[color=myplotblue,very thick,densely dashed] [samples=2] {1.675};
\addlegendentry{No refinement}
\end{axis}
\end{tikzpicture}
\centering
\caption{Comparison of sampling methods in terms of Wasserstein--1 divergence $W_1$ against unbiased samples of $p(x)$ obtained via rejection sampling, for a randomly initialized \FBM{} with $N=20$. We set $K = 4N$ and report results from ULA and MALA with optimized hyperparameters ($\epsilon_t = 10^{-5}$ and $\epsilon_t = 8 \times 10^{-5}$ respectively).}
\label{fig:sampling_comparison}
\end{figure}
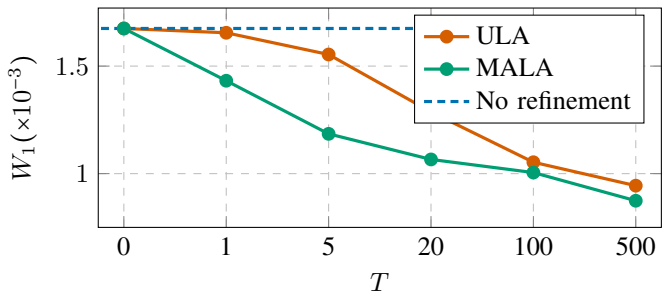

We experimentally validate our algorithm by studying the effect of the number of discretization points $K$, and the effect of the number of sampling steps $T$ on the ULA and MALA refinement.

\subsection{Varying $K$}
\cref{fig:empirical-distributions-2} gives an illustration of the sampling distribution $q(x)$ vs. the model density $p(x)$ as $K$ increases. A qualitative comparison of the histograms with and without ULA/MALA refinement is provided in \cref{fig:empirical_distributions} in the appendix of the pre-print. \cref{fig:approximation-convergence} shows the Kullback--Leibler divergence $\DKL(p, q)$ as a function of $K$ for varying number of frequency terms $N = \{50, 100, 200\}$ (left pane) and varying choices of the interpolation kernel $w_D$ (right pane). For each configuration, we show $20$ different trials, each testing an \FBM{} with randomly sampled coefficients. We estimate the divergences using Monte Carlo (MC) sampling with unbiased samples from $p(x)$ obtained via rejection sampling. As $K$ increases, divergences drop monotonically. In the left pane we confirm that since a larger $N$ implies a more complicated distribution, $K$ must also be larger to achieve the same divergence. In the right pane we observe that the triangular kernel $w_1$ outperforms other choices empirically. See \cref{fig:kernel-interpolation} in the pre-print for a qualitative comparison.

\subsection{Varying $T$}
We evaluate the effect of the ULA and MALA refinements as described in \emph{Step 3}. Since the density of the sampling distribution resulting from running MCMC methods for a fixed number of steps $T$ is generally not known in closed form, we cannot use MC to estimate divergences as above. Instead, we compute the Wasserstein divergence $W_1$ between empirical distributions sampled from $q(x)$, and empirical distributions sampled from $p(x)$ via rejection sampling.

 \cref{fig:sampling_comparison} visualizes the Wasserstein--1 divergence of the sampling distribution as a function of the number of MCMC sampling steps $T$. We observe that both ULA and MALA produce increasingly better samples as the number of sampling steps $T$ increases and lead to better approximations in comparison to DAAS without refinement. However, this accuracy comes at a significantly larger computational cost as demonstrated in \cref{tab:results}. Without refinement, DAAS only requires $K$ FBM evaluations, independently of the number of samples to be drawn. MCMC methods further need to evaluate the \FBM{} model for each sample and sampling step.

\begin{table}
\centering
\begin{tabular}{@{}lc@{}}
\toprule
\textbf{Method}  & Number of \FBM{} Evals  \\
\midrule
Rejection Sampling & $5\times 10^7$ \\
ULA (T=20)& $4\times 10^7 + 50$  \\
MALA (T=20) &  $8\times 10^7 + 50$ \\
Triangular  &  $50$ \\
\bottomrule
\end{tabular}
\caption{Computational cost of various methods in terms of number of \FBM{} model evaluations to draw $10^6$ samples, assuming $N=10$ and $K=50$.}
\label{tab:results}
\end{table}

\section{Conclusion}
\label{sec:conclusion}
This paper introduces a general and flexible approximate sampling algorithm designed for Fourier Basis Density Model (\FBM{})\cite{delafuente2024fourierbasisdensitymodel}. The algorithm leverages the mathematical properties of \FBM{}, in particular its band-limitedness, to achieve computational efficiency. We perform a systematic evaluation of several proposed sampling methodologies, highlighting trade-offs between computational costs associated with sampling and the resulting accuracy of the generated samples. Furthermore, we present theoretical properties and bounds, both for the previously proposed \FBM{} model and for the proposed sampling algorithm. Our method enables efficient sampling from the \FBM{}, opening the door for practical applications.

\clearpage
\printbibliography

\clearpage
\appendices

\begin{figure*}[t]
\centering
\begin{subfigure}{0.48\textwidth} 
    \centering
    \includegraphics[width=\textwidth]{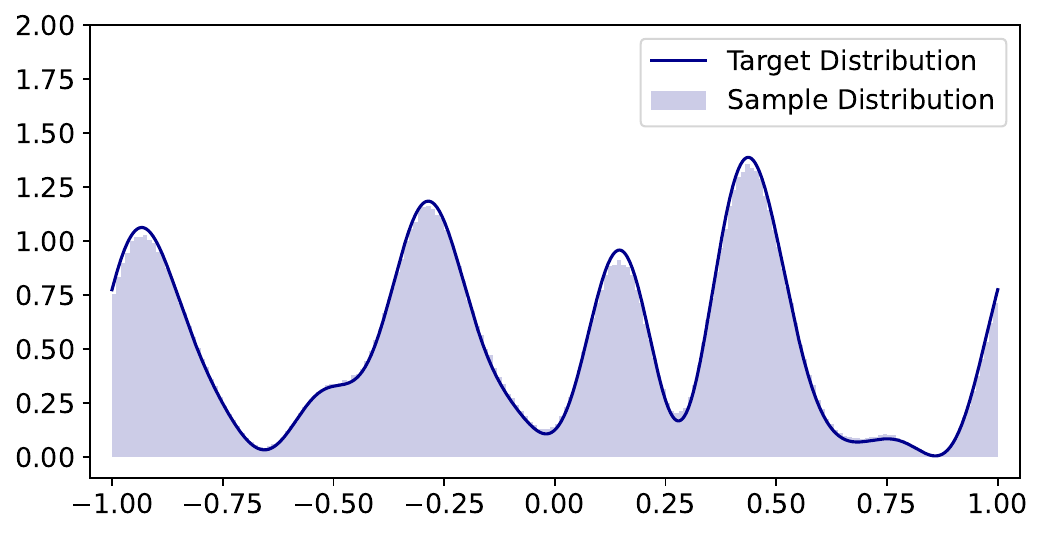}
    \caption{DAAS-Unadjusted Langevin Dynamics (ULA)}
\end{subfigure}\hfill 
\begin{subfigure}{0.48\textwidth}
    \centering
    \includegraphics[width=\textwidth]{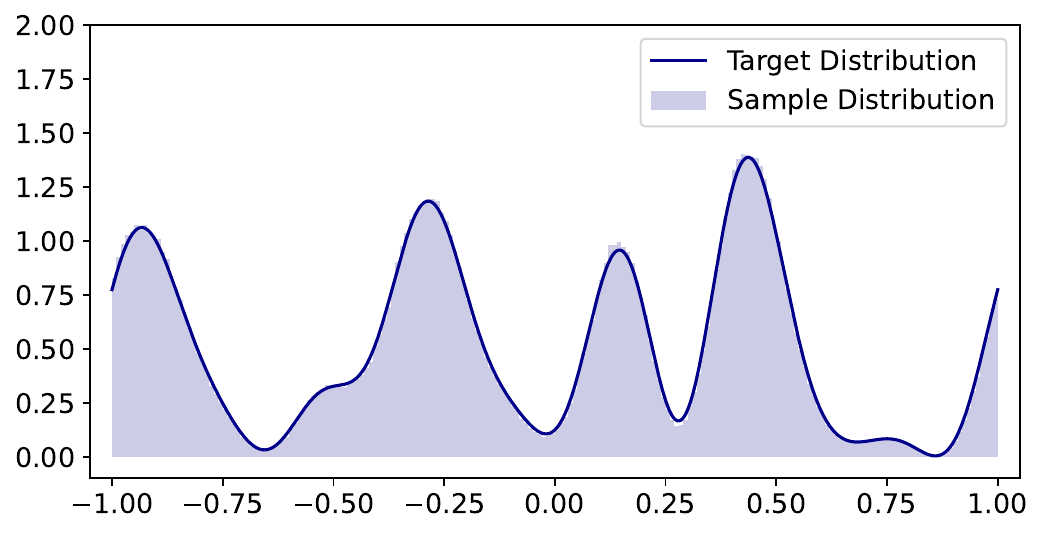}
    \caption{DAAS-Metropolis-adjusted Langevin Algorithm (MALA)}
\end{subfigure}

\begin{subfigure}{0.48\textwidth}
    \centering
    \includegraphics[width=\textwidth]{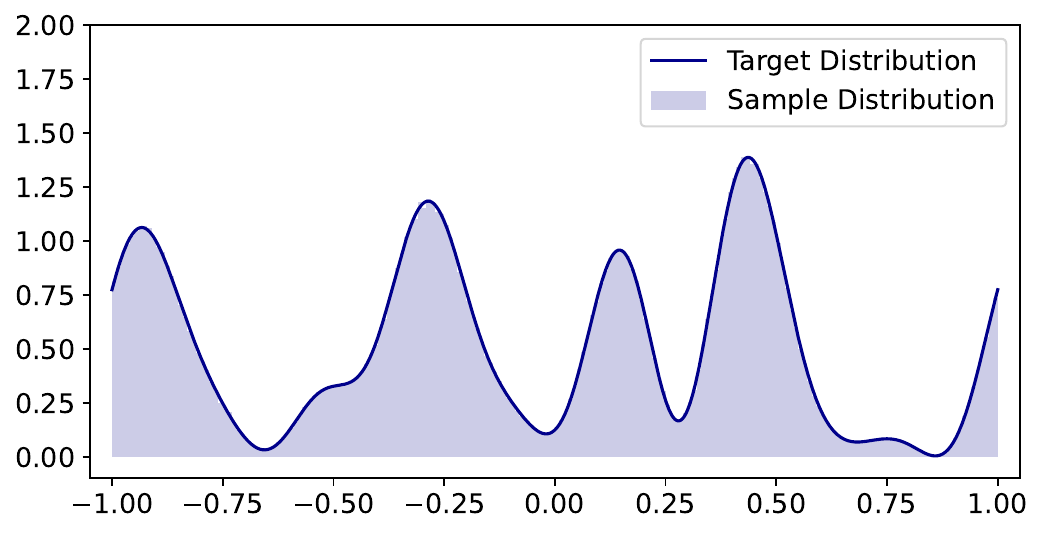} 
    \caption{Rejection Sampling}
\end{subfigure}\hfill
\begin{subfigure}{0.48\textwidth}
    \centering
    \includegraphics[width=\textwidth]{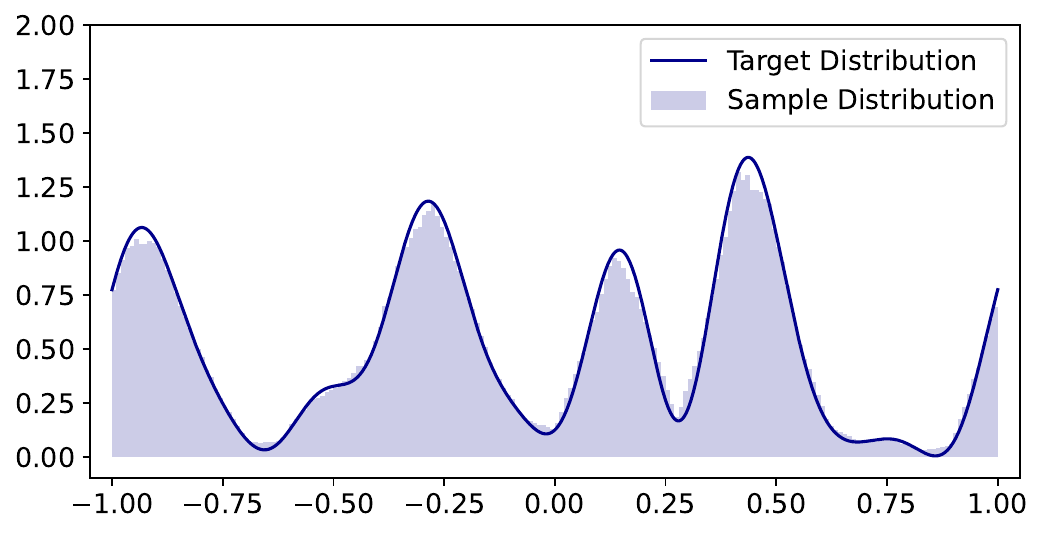} 
    \caption{\textbf{Discretized Approximate Ancestral Sampling}}
\end{subfigure}
\caption{Visual comparison of empirical distributions by different sampling methods for an arbitrary multi-modal \FBM{} density with $N=10$ frequency terms and $K=30$ sampling points. We observe how the empirical distributions of $10^6$ samples for each method 
differ to capture a few local minima/maxima present on the \FBM{}  distribution.}
\label{fig:empirical_distributions}
\end{figure*}

\section{Positivity}
\label{sec:prop1_proof_app}
\begin{prop}
Let $\{a_k\}_{k=0}^N$ be a sequence of complex numbers. We define the sequence $\{c_n\}_{n=0}^N$ by
\begin{equation}
c_n = \sum_{k=0}^{N-n} a_k a_{k+n}^*, \quad n = 0, 1, \dots, N.
\end{equation}
Then, for $f$ as defined in \eqref{eq:fourier_series}, we have $f(x) \ge 0$, $ \forall x \in [-1, 1)$.
\end{prop}

\begin{IEEEproof}
Let us consider the function
\begin{equation}
A(x) = \sum_{k=0}^N a_k \exp(-i k \pi x).
\end{equation}
Then,
\begin{align}
|A(x)|^2 &= A(x) A^*(x) \\
&= \left( \sum_{k=0}^N a_k \exp(-i k \pi x) \right) \left( \sum_{j=0}^N a_j^* \exp(i j \pi x) \right) \\
&= \sum_{k=0}^N \sum_{j=0}^N a_k a_j^* \exp(i(j-k)\pi x) \\
&= \sum_{n=-N}^N \left( \sum_{k=0}^{N-n} a_k a_{k+n}^* \right) \exp(i n \pi x) = f(x).
\end{align}
Since $|A(x)|^2 \ge 0$ for all $x$, we have $f(x) \ge 0$ for all $x \in [-1, 1)$.
\end{IEEEproof}

\section{Proof of \cref{prop:average_value}}
\label{sec:prop5_proof_app}
\begin{IEEEproof}
\begin{align}
\sum_{k=0}^{K-1} p(x_k) &= \sum_{k=0}^{K-1} \left( \frac{1}{2} + \sum_{n=1}^N \Re \left\{ \frac{c_n}{c_0} e^{i n \pi x_k} \right\} \right) \\
&= \frac{K}{2} + \sum_{n=1}^N \Re \left\{ \frac{c_n}{c_0} \sum_{k=0}^{K-1} e^{i n \pi x_k} \right\} \\ 
&= \frac{K}{2} + \frac{1}{c_0} \sum_{n=1}^N   \Re \left\{ c_n  \sum_{k=0}^{K-1} \left(  e^{i 2 \pi  n/K} \right)^k  \right\} \\
&= \frac{K}{2}  + \frac{1}{c_0} \sum_{n=1}^N   \Re \left\{ c_n  \sum_{k=0}^{K-1} \frac{e^{ 2 \pi i } - 1 }{e^{2 \pi i /K } - 1} \right\} = \frac{K}{2}. 
\end{align}
\end{IEEEproof}
This result originally occurred as a lemma in \cite{gillman2024fourierheadhelpinglarge}.

\section{Proof of \cref{prop:piecewise_linear_approximation}}
\label{sec:prop9_proof_app}
\begin{IEEEproof}
First, we verify that $q(x)$ is indeed a valid distribution. Since it is non-negative by definition, we only need to verify that it integrates to $1$ over the domain $[-1, 1)$.
\begin{align}
    \int_{-1}^{1} q(x) dx &= \int_{-1}^{1} \sum_{k=0}^{K-1} p[k] \, \tfrac K 2 w_1\bigl(\tfrac K 2 (x - x_k)\bigr) \, dx \\
    &= \sum_{k=0}^{K-1} p[k] \int_{-1}^1 \tfrac K 2 w_1\bigl(\tfrac K 2 (x - x_k)\bigr) \, dx \\
    & = \sum_{k=0}^{K-1} p[k] = 1
\end{align}
The last step uses the result from \cref{prop:average_value}. Further, since the individual triangle kernels wrap around the boundaries of the domain $[-1, 1)$, they integrate to $1$ on the full domain.

Next, note that within the interval $[x_k, x_{k+1}]$, the only non-zero terms in $q(x)$ are the triangular distributions shifted to $x_k$ and $x_{k+1}$. All other triangular distributions are zero in this interval. Therefore, we have within $[x_k, x_{k+1}]$ that:
\begin{multline}
q(x) = p[k] \, \tfrac K 2 w_1\bigl(\tfrac K 2 (x - x_k)\bigr) \\
+ p[k+1] \, \tfrac K 2 w_1\bigl(\tfrac K 2 (x - x_{k+1})\bigr).
\end{multline}
Substituting eq. \eqref{eq:triangle} and simplifying, we have: 
\begin{equation}
q(x) = \frac { p(x_k) \, (x_{k+1} - x) + p(x_{k+1}) \, (x - x_k) } { x_{k+1} - x_k }
\end{equation} 
This expression corresponds to a linear function of $x$ within the interval $[x_k, x_{k+1}]$. Therefore, the compound distribution $q(x)$ is a piecewise linear function, where each piece is a linear interpolation between the values of the original distribution $p(x)$ at the $K$ equally spaced points $x_k$.
\end{IEEEproof}

\section{Properties of the \FBM}

\subsection{Scale invariance}
\begin{property}
The \FBM{} density $p(x)$ is invariant to the scale of the sequence $\{a_k\}_{k=0}^N$.
\end{property}

\begin{IEEEproof}
If we scale the sequence $\{a_k\}_{k=0}^N$ by a factor $\alpha \in \mathbb{C}$, i.e.,  $a'_k = \alpha a_k$, then the new coefficients $c'_n$ become:
\begin{align}
    c'_n &= \sum_{k=0}^{N-n} a'_k (a'_{k+n})^*  \\
    &= \sum_{k=0}^{N-n} \alpha a_k (\alpha a_{k+n})^* \\
    &= |\alpha|^2 \sum_{k=0}^{N-n} a_k a_{k+n}^* \\
    &= |\alpha|^2 c_n
\end{align}
Consequently, the density function $p(x)$ remains unchanged, as the factors of $|\alpha|^2$ cancel out in the ratio $c_n/c_0$. Therefore only the relative magnitudes and phases of $a_k$ affect the shape of $p(x)$.
\end{IEEEproof}

\subsection{Finite zeros}
\begin{property}
The \FBM{} density $p(x)$ defined over $[-1, 1)$ has a finite set of zeros, upper bounded by $2N$, with $N$ being the number of frequency terms.
\end{property}

\begin{IEEEproof}
By substituting $z \equiv e^{i\pi x}$ in the $p(x)$ as defined in \eqref{eq:model_p}, we obtain a polynomial in terms of the variable $z$ of degree at most $2N$, where $N$ is the number of frequency terms of $p(x)$.  Let's rewrite $p(x)$ as follows,
\begin{align}
    p(x) &= \frac{1}{2} + \frac{1}{2}\sum_{n=1}^N \left( \frac{c_n^*}{c_0} z^n + \frac{c_n}{c_0} z^{-n} \right) \\
    2 c_0 p(x) &= c_0 + \sum_{n=1}^N ({c_n^*} z^n + c_n z^{-n}) \\
    2 c_0 z^N p(x) &= c_0 z^N + \sum_{n=1}^N ({c_n^*} z^{n+N} + c_n z^{N-n}) 
\end{align}
Let $P(z) = 2 c_0 z^N p(x)$, if we set $p(x) = 0$, then $P(z) = 0$. $P(z)$ is a polynomial in $z$ of degree $2N$. By the Fundamental Theorem of Algebra, a polynomial of degree $2N$ can have at most $2N$ roots. Thus $p(x)$ has at most $2N$ distinct real roots. 
\end{IEEEproof}

\subsection{Minimum Zeros Spacing}
\begin{property}
For any positive trigonometric polynomial $p(x)$ defined over $[-1, 1)$ with $N$ frequency terms, any two distinct real zeros $x_1 < x_2$ of $p$ satisfy $x_2 - x_1 \ge \frac{2 \ p(x^*)}{\pi N  \Vert p\Vert_{\infty}} $, such that $p$ attains a local maximum at $x_1 < x^* < x_2$.
\end{property}
\begin{IEEEproof}
Assume two distinct zeros of $p$ at points $x_1 < x_2$ of $p$  are separated by $\Delta x=x_2-x_1$. Since $p(x) \geq 0$ is continuous and differentiable, and $p(x_1) = p(x_2) = 0$, it must attain a local maximum at $x^*$ where $x_1 <x^* < x_2$. Then, 
\begin{align}
    \int_{x_1}^{x_2} | p'(x) | dx &= \int_{x_1}^{x^*} | p'(x) | dx +  \int_{x^*}^{x_2} | p'(x) | dx  \\
    &\geq \Bigg | \int_{x_1}^{x^*} p'(x) dx  \Bigg | + \Bigg| \int_{x^*}^{x_2}  p'(x)  dx  \Bigg | \\
    &= | p(x^*) - p(x_1) | + | p(x^*) - p(x_2) | \\
    &= 2 p(x^*).
\end{align}
Thus, 
\begin{align}
    2 \max_{[x_1, x_2]}  p(x)  \leq \int_{x_1}^{x_2} | p'(x) | dx \leq ( x_2 - x_1 )  \max_{[x_1, x_2]}| p'(x) | 
\end{align}
We know by Bernstein's inequality \cite{queffélec2019bernsteinsinequalitypolynomials} for a trigonometric polynomial $p(x)$ of degree $N$ $ \left( \sum_{n=-N}^N c_n e^{\pi i n x} \right) $,
\begin{align}
    \|p'(x)\|_\infty \le \pi N\|p(x)\|_\infty.
\end{align}
Combining previous results, 
\begin{align}
    2 \max_{[x_1, x_2]} | p(x) | \leq  ( x_2 - x_1 ) \pi N \|p(x)\|_\infty.
\end{align}
Since $\|p(x)\|_\infty \geq \frac12$, we can conclude 
\begin{align}
x_2 - x_1 \geq \frac{2 \max_{[x_1, x_2]} p(x)}{\pi N \|p(x)\|_\infty}.
\end{align}
In particular, with $\max_{[x_1, x_2]} p(x) = \|p(x)\|_\infty$, 
\begin{align}
    x_2 - x_1 \geq \frac{2}{\pi N}
\end{align}
\end{IEEEproof}

\subsection{Coefficient decay}
\begin{property}
Let $f(x)$ be as specified in \eqref{truncatedeq}. In general, if $f$ is $k$-times differentiable, there exists a constant $C > 0$ such that:
\begin{align}
    |c_n| \le \frac{C}{n^k},  \quad \text{for } n \ge 1 
\end{align}
\end{property}

\begin{IEEEproof}
Let us recall \eqref{eq:coefficients} and integrate by parts,
\begin{align}
    c_n &= \frac{1}{2} \left[ \frac{f(x) e^{-\pi i n x }}{-i n \pi } \right]_{-1}^1 - \frac{1}{2} \int_{-1}^1 f'(x) \frac{e^{- \pi i n  x }}{-i n \pi } dx \\
    &= \frac{i}{2 n \pi} \int_{-1}^1 f'(x) e^{-i n \pi x } dx 
\end{align}  

We can repeat this process $k$ times since $f(x)$ is infinitely differentiable. After $k$ integration by parts we get:
\begin{align}
    c_n = \left(  \frac{i}{n \pi} \right)^k \frac{1}{2} \int_{-1}^1 f^{(k)}(x) e^{- \pi i n  x } dx
\end{align}
Then, taking the magnitude of $c_n$,
\begin{align}
    |c_n| = \frac{1}{2 (n \pi)^k} \left| \int_{-1}^1 f^{(k)}(x) e^{- \pi i n  x } dx \right|
\end{align}

Since $f^{(k)}(x)$ is continuous on the closed interval $[-1, 1)$, it is bounded. Then,
\begin{align}
 |c_n| \le \frac{1}{2 (n \pi)^k} \int_{-1}^1 |f^{(k)}(x)| dx 
\end{align}
By applying, Bernstein inequality for trigonometric polynomials of degree $N$,
\begin{align}
    |c_n| &\le \frac{1}{2 (n \pi)^k} \int_{-1}^1 (\pi N)^k \|f(x)\|_\infty dx  = \left( \frac{N}{n} \right)^k \|f(x)\|_\infty  . 
\end{align}
\end{IEEEproof}

\subsection{Zeroth coefficient bound}
\begin{property}
\label{prop:czero_bound}
$ |c_n| \le c_0 , \quad n \in \{1,\cdots, N \}$
\end{property}

\begin{IEEEproof}
We are given that
\begin{equation}
c_n = \sum_{k=0}^{N-n} a_k a_{k+n}^*, \quad n = 0, 1, \dots, N.
\end{equation}
By Cauchy--Schwarz inequality, 
\begin{align}
|c_n|^2 &= \left| \sum_{k=0}^{N-n} a_k a_{k+n}^* \right|^2 \\
&\le \left( \sum_{k=0}^{N-n} |a_k|^2 \right) \left( \sum_{k=0}^{N-n} |a_{k+n}|^2 \right) \\
&= \left( \sum_{k=0}^{N-n} |a_k|^2 \right) \left( \sum_{k=n}^N |a_k|^2 \right).
\end{align}
Now, observe that
\begin{equation}
    c_0 = \sum_{k=0}^N a_k a_k^* = \sum_{k=0}^N |a_k|^2.
\end{equation}
Since $0 \le n \le N$, we have $0 \le N-n \le N$, and thus
\begin{equation}
    \sum_{k=0}^{N-n} |a_k|^2 \le \sum_{k=0}^N |a_k|^2 = c_0, \quad  \sum_{k=n}^N |a_k|^2 \le \sum_{k=0}^N |a_k|^2 = c_0.
\end{equation} 
Therefore,
\begin{align}
 |c_n|^2 & \le  \left( \sum_{k=0}^{N-n} |a_k|^2 \right) \left( \sum_{k=n}^N |a_k|^2 \right) \\
 &\le c_0 \cdot c_0 = c_0^2.
\end{align}
Since $c_0 > 0$, $|c_n| \le c_0$ for $n = 1, 2, \dots, N$.
\end{IEEEproof}

\subsection{Bounded first and second derivatives}
\begin{property}
For any non-constant \FBM{} density $p(x)$, we have that  $|p'(x)|$ and $|p''(x)|$ are bounded.
\end{property}

\begin{IEEEproof}
Given that the density defined in \eqref{eq:model_p} is $k$ times continuously differentiable ($k>3)$ we have:
\begin{align}
p'(x) &= \sum_{n=1}^N \Re \left( \frac{c_n}{c_0} (in\pi) \exp(\pi i n  x) \right), \\
p''(x) &= \sum_{n=1}^N \Re \left( \frac{c_n}{c_0} (in\pi)^2 \exp(\pi i n x) \right).
\end{align}
By using Property \ref{prop:czero_bound},
\begin{align}
|p'(x)| & \le \sum_{n=1}^N n\pi \left| \frac{c_n}{c_0} \right| \le \sum_{n=1}^N n\pi = \frac{N(N+1)}{2}\pi
\end{align}
and
\begin{align}
|p''(x)| &\le \sum_{n=1}^N n^2\pi^2 \left|\frac{c_n}{c_0}\right|  \le \sum_{n=1}^N n^2\pi^2 \\
&= \frac{N(N+1)(2N+1)}{6}\pi^2.
\end{align}
\end{IEEEproof}

\subsection{Linear interpolation bound}
\begin{property}
\label{prop:linear_interpolation}
For any non-constant \FBM{} density $p(x)$ defined on $[-1, 1)$, let $q(x)$ be the piecewise linear interpolation of $p$ using points $x_j = -1 + 2j/K$, for $j=0, 1, \dots, K-1$. Then, for any $x \in [-1, 1)$, the error of the interpolation is bounded by:
\begin{align}
|p(x) - q(x)| \le \frac{\pi^2 N ( N + 1 ) (2N + 1) }{12 K^2}
\end{align}
\end{property}

\begin{IEEEproof}
Based on standard result for interpolation methods \cite{ralston2001first}, if $p(x)$ is a function defined on $[-1, 1)$ with a continuous second derivative, such that $|p''(x)| \le M$ for all $x \in [-1, 1)$, and $q(x)$ is a piecewise linear interpolation at $K$ equally spaced points within $[-1, 1)$ (same setting as \cref{prop:piecewise_linear_approximation}), we have  
\begin{align}
    |p(x) - q(x)| \le  \frac{M}{2K^2}
\end{align}
For our particular case, 
\begin{align}
    |p(x) - q(x)| \le  \frac{\pi^2 N(N+1)(2N+1)}{12K^2}
\end{align}
the result holds for all $x \in [-1, 1)$.
\end{IEEEproof}

\section{Proof of \cref{thm:divergence_bounds}}
\label{sec:thm1_proof_app}
\begin{IEEEproof}
By definition, and using the linear interpolation error bound (\cref{prop:linear_interpolation}), 
\begin{align}
    \DTV(p,q) &= \frac12 \int_{-1}^1 \vert p(x) - q(x) \vert \, dx \\
        &\leq \frac12 \int_{-1}^1 \frac{\pi^2 N(N+1)(2N+1)}{12K^2} \, dx \\
        &= \frac{\pi^2 N(N+1)(2N+1)}{12K^2}
\end{align}
This proves \eqref{eq:tvd_bound}. With a similar approach, and using the integral triangle inequality we prove an upper-bound for the Wasserstein divergence $W_1$,
\begin{align}
    \DWO(p,q) &= \int_{-1}^1  \left| \int_{-1}^x p(t) dt -  \int_{-1}^x q(t) dt \right| \, dx \\
    &= \int_{-1}^1  \left| \int_{-1}^x (p(t) - q(t)) dt \right| \, dx \\
    &\leq \int_{-1}^1  \int_{-1}^x \vert p(t) - q(t) \vert \, dt \, dx \\
    & \leq  \int_{-1}^1  \int_{-1}^x \frac{\pi^2 N(N+1)(2N+1)}{12K^2} \, dt \, dx \\
    &= \frac{\pi^2 N(N+1)(2N+1)}{6K^2}.
\end{align} 
This proves \eqref{eq:wasser_bound}.
\end{IEEEproof}

\section{B-splines interpolation}
\begin{figure}
\centering 
\begin{tikzpicture}
  \def\knotVectorOne{-0.5, -0.5, -0.5, 0.5, 0.5, 0.5}
  \def\knotVectorTwo{-1, -1, -1, -1, 0, 0, 0, 0}
  \def\knotVectorThree{-1.5, -1.5, -1.5, -1.5, -0.5, 0.5, 0.5, 0.5, 0.5, 1.5}
  \def\knotVectorFour{-2, -2, -1, -1, -1, -1, 0, 0, 0, 0, 1, 1, 2}

    \def\bSplineScale{1}

  \begin{axis}[
    xmin=-2, xmax=2,
    ymin=0, ymax=1.1,
    axis lines=center,
    axis line style={-},
    xtick={-2,-1.5,-1,-0.5,0,0.5,1,1.5,2},
    ytick={0,0.5,1},
    xlabel={},
    ylabel={},
    clip=false, 
    width=\linewidth, 
    height=4cm,
    legend style={font=\small},
  ]
  \addplot[myplotgreen, very thick] coordinates {(-0.5, 0) (-0.5, \bSplineScale)};
  \addplot[myplotgreen, very thick, forget plot] coordinates {(0.5, 0) (0.5, \bSplineScale)};
  \addplot[domain=-0.5:0.5, samples=100, myplotgreen, very thick, forget plot]
    { (x >= -0.5 && x <= 0.5) ? \bSplineScale : 0 }; 

  \addplot[domain=-1:0, samples=100, myplotred, very thick, densely dashdotted]
    {(x >= -1 && x <= 0) ? \bSplineScale*(x + 1) : 0 };
  \addplot[domain=0:1, samples=100, myplotred, very thick, densely dashdotted, forget plot]
   { (x >= 0 && x <= 1) ?  \bSplineScale*(-x + 1) : 0 };  

  \addplot[domain=-1.5:-0.5, samples=100, myplotblue, very thick, dashed]
    { (x >= -1.5 && x <= -0.5) ? \bSplineScale*0.5*(x+1.5)*(x+1.5) : 0};
    \addplot[domain=-0.5:0.5, samples=100, myplotblue, very thick, dashed, forget plot]
    { (x >= -0.5 && x <= 0.5) ? \bSplineScale*(-x*x + 0.75) : 0}; 
    \addplot[domain=0.5:1.5, samples=100, myplotblue, very thick, dashed, forget plot]
    { (x >= 0.5 && x <= 1.5) ? \bSplineScale*0.5*(1.5-x)*(1.5-x) : 0 };

  \addlegendentry{$w_0(x)$}
  \addlegendentry{$w_1(x)$}
  \addlegendentry{$w_2(x)$}
  \end{axis}
\end{tikzpicture}
\caption{First four B-spline functions $w_D$.} 
\label{fig:bsplines} 
\end{figure}
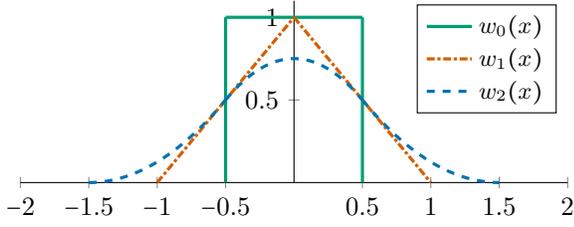

The approach discussed in Algorithm~\ref{alg:daas} works in principle for arbitrary B-spline interpolation filters $w$ (as shown in \cref{fig:bsplines}) which correspond to the Irwin--Hall distributions \cite{hall1927distribution}.

\begin{figure}
  \centering 
  \begin{subfigure}[b]{0.85\linewidth}
    \centering
    \includegraphics[width=\linewidth, keepaspectratio]{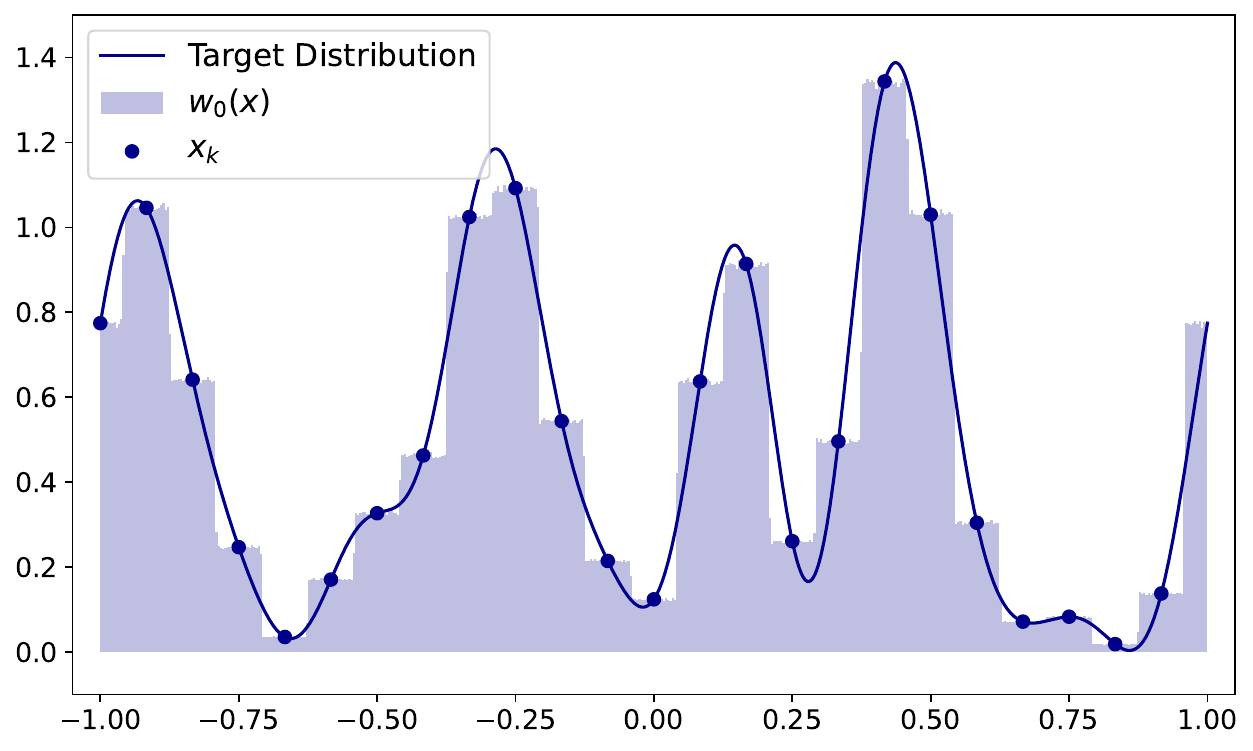}
  \end{subfigure}
  \begin{subfigure}[b]{0.85\linewidth} 
    \centering
    \includegraphics[width=\linewidth, keepaspectratio]{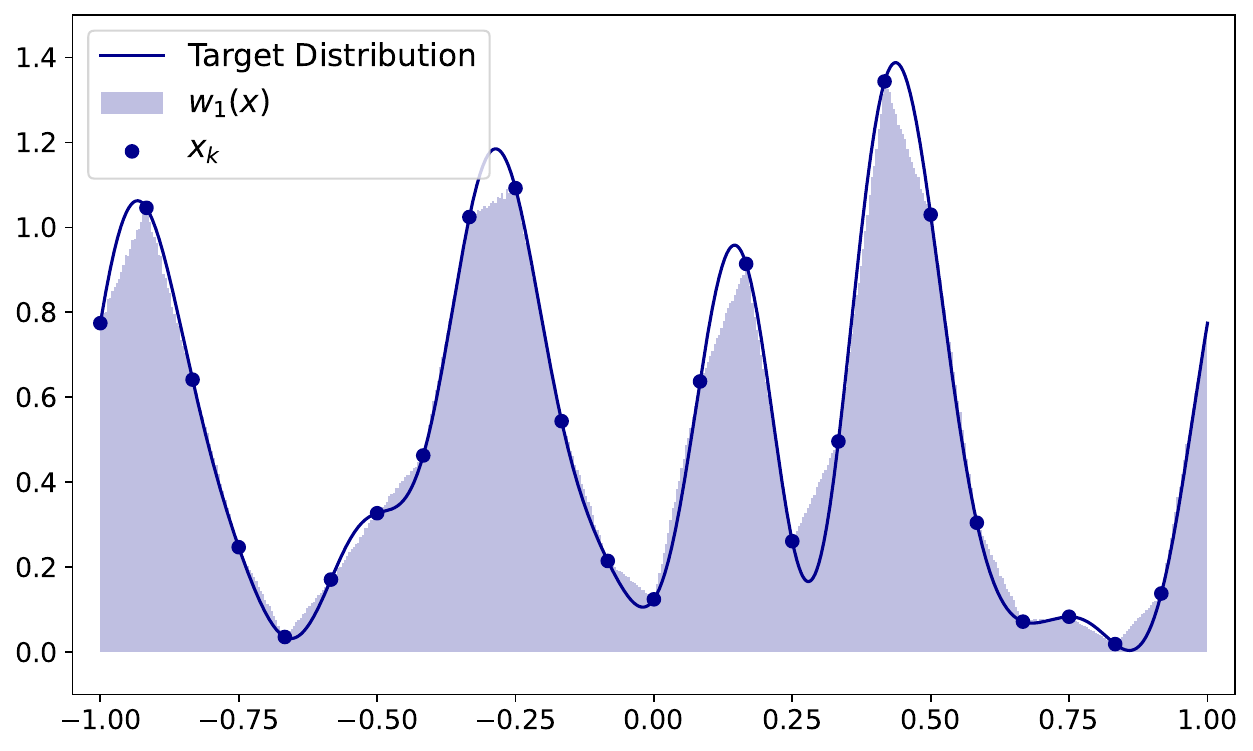}
  \end{subfigure}
  \begin{subfigure}[b]{0.85\linewidth}
    \centering
    \includegraphics[width=\linewidth, keepaspectratio]{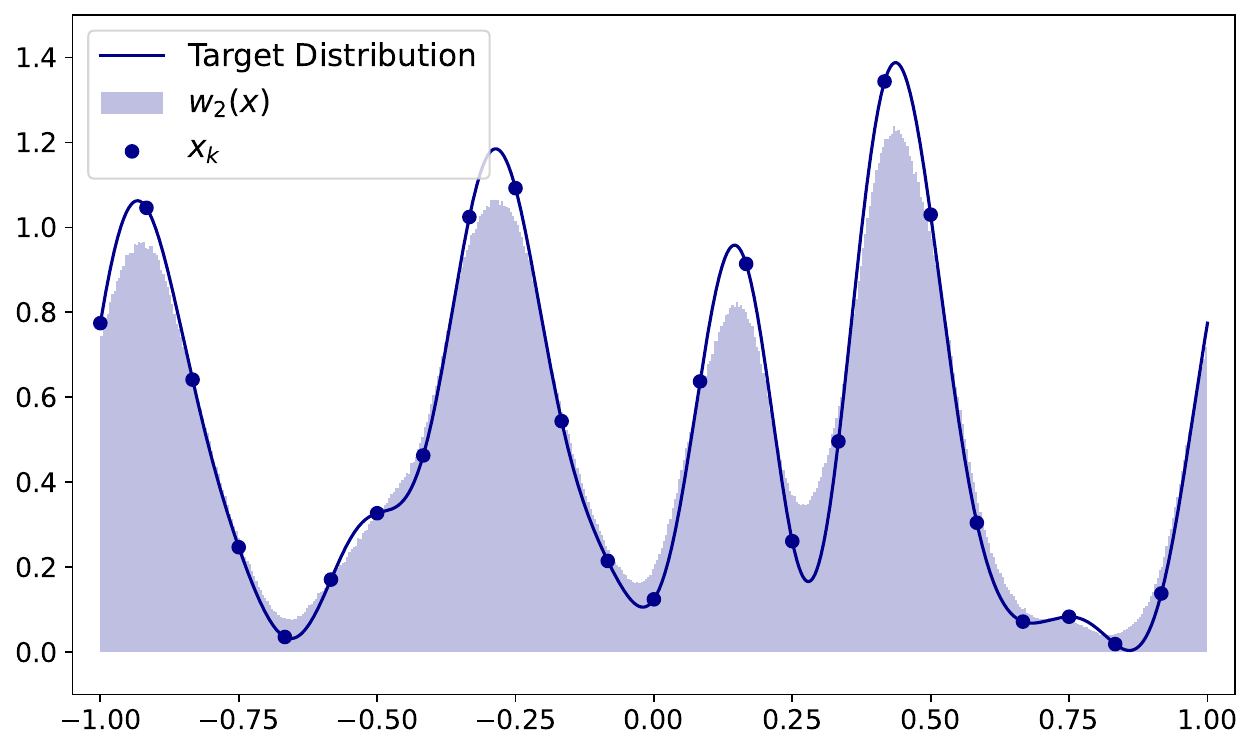}
  \end{subfigure}
  \caption{Visualization of uniform $w_0$, triangular $w_1$, and quadratic $w_2$ filters with respect to target distribution.}
  \label{fig:kernel-interpolation}
\end{figure}

Each interpolation filter $w$ smoothes out the target distribution as seen in \cref{fig:kernel-interpolation} for uniform, triangular and piecewise quadratic B-spline filters.

\section{ULA and MALA details}
\label{sec:ula_mala_details}
ULA corresponds to a discretization of the Langevin stochastic differential equation, which uses the score function to guide the sampling as following, 
\begin{align}
    x_{t+1} = x_t + \epsilon_t \nabla \log p(x) + \sqrt{2\epsilon_t} z , \quad z_t \sim \mathcal{N}(0, I)
\end{align}
 where $x_0 \sim p[k]$ is a sample from the ancestor distribution and $\epsilon_t$ is the time dependent step size of the method. The choice of step size plays a critical role in the convergence of the Langevin-Dynamics based algorithms \cite{teh2015consistencyfluctuationsstochasticgradient}. In this paper, we set $0 < \epsilon_t \ll 1$ as a constant or decaying with the iterations as $\epsilon_t = \epsilon_0 / (t + 1) $. Note that the discretization introduces bias in the sampling, with higher bias for higher step sizes in general. Therefore, small or decaying step sizes tend to perform well in practice.

MALA views a step of ULA as proposing a sample from a proposal distribution. To remove the bias due to the discretization, it incorporates the following Metropolis-Hastings based accept-reject step leading to unbiased samples once the chain converges.
 \begin{align}
\text{Acceptance: } & \alpha(x', x_t) = \min\left(1, \frac{p(x') r(x_t | x')}{p(x_t) r(x' | x_t)}\right) \\
r(x' | x_t) &\propto \exp \left( \frac{-\| x' - x_t - \epsilon_t \nabla \log p(x_t)\|_2^2}{4\epsilon_t}  \right)  \\
\text{Update: } &
\begin{cases}
x_{t+1} = x' & \text{if } u \le \alpha, \quad u \sim \mathcal{U}(0, 1) \\
x_{t+1} = x_t & \text{otherwise}
\end{cases}
\end{align}
Here $x_t$ is the current state and $x'$ is the proposal being considered. $\alpha$ is the acceptance probability and $r$ denotes the proposal distribution.
Although theoretical guarantees are well-known for ULA and MALA, it is worth noting that for our case of circular distributions we need to adjust for warping operator at each iteration.  

\end{document}